\documentstyle[prl,aps,epsf,eqsecnum]{revtex}
\tightenlines
\begin{document}
\title{Symmetry in Order-Disorder Changes of Molecular Clusters}
\author{Ana Proykova$^\dagger$, Dessislava Nikolova$^\dagger$,
and R.Stephen Berry$^*$}
\address{$^{\dagger}$University of Sofia, Faculty of Physics,
5 James Bourchier Blvd., Sofia-1126, Bulgaria}
\address{$^*$The University of Chicago, Department of Chemistry
Chicago, IL 60637, USA}
\maketitle
\firstfigfalse
\begin{abstract}
The dynamic orientational order-disorder transition of clusters consisting
of octahedral $AF_6$ molecules is formulated in terms of 
symmetry-adapted rotator functions.
The transition from a higher-temperature 
body-centered-cubic phase whose
molecules are orientationally disordered at their sites to lower-temperature, 
monoclinic, orientationally-ordered phase is a two-step process: first, at 
temperatures well below the limit of stability for the liquid, a transition
occurs to a
partially ordered monoclinic phase driven by the rotational-vibrational
coupling. This transition has two local minima in the free energy,
and hence behaves like a finite-system counterpart of a first-order 
transition.  Further lowering of the temperature initiates another
transition, to an orientationally-ordered base-centered monoclinic structure. 
This last transition is dominated by rotational-rotational interaction and 
is found from simulations to be continuous. The temperature of 
this transition predicted 
by the analytic theory presented here for a 59-molecule cluster of $TeF_6$, 27K, is in 
good agreement with the ~30K result of canonical Monte Carlo calculations.

\end{abstract}

\newpage

Below the freezing point, the plastic phases of molecular substances are known to form highly symmetrical lattices with orientationally disordered molecules \cite{TIM61}. As the temperature of the disordered substance is  still lowered, a phase transition to a more orientationally-ordered phase takes place. The new phase might have a crystal structure of lower symmetry, characterized by long-range 
orientational order. This is called displacive-ordering transition. Similar structural transformations have been also detected in small, free clusters consisting of rigid octahedral molecules 
$SF_6$ \cite{FAR83} and $TeF_6$ \cite{BAR87,BER99} and have been studied by numerical simulations \cite{FAR86,BAR91,BER97,BER98,BER98A} and
experimentally \cite{RAO89,SCH97}. The change between phases 
may be continuous or discontinuous depending on the interaction potential. However, simulations of 
clusters always produce smooth behavior in the sense that clusters in different phases coexist 
over a range of temperatures and pressures,  regardless of the order of the transition in the
thermodynamic, large-N limit.  In many cases, the
crystal forms observed depend strongly on the conditions of production, thus 
leading to different interpretations of what should be the final structures
of the clusters at low temperatures. Structures such as triclinic, monoclinic and hexagonal
were reported for the same material \cite{BAR91}. In our previous simulations, we have shown that
the finite size of clusters causes rotational \cite{BER97} and strong
surface effects \cite{BER99A}, appearance of vibrational-rotational coupling, none of which are expected in bulk systems of the same symmetry. 

Vibrational-rotational coupling was considered qualitatively
to explain the two-step process of ordering in some clusters of octahedral molecules, $O_h$ symmetry, as
the temperature decreases \cite{BER99,BER98}. Below the melting temperature,
those clusters assume a body-centered
cubic structure $\it bcc\/$,  $O_h$ symmetry, and orientational disorder of 
the molecules \cite{FAR83,FAR86,BAR87}.
Simulations show \cite{BER99B} that this transition 
involves passage between two forms at different local minima in the 
free energy, and hence is the apparent counterpart of a first-order 
transition.  Whether these two minima remain apart or converge as 
$N \rightarrow \infty$ is an unsettled question \cite{FAR83,BAR87}.
A temperature decrease drives 
a phase transition from the orientationally disordered {\it bcc\/} structure
to an orientationally ordered monoclinic phase, $C_{2h}$.
Since no normal mode of $O_h $ carries the group directly to $C_{2h}$ this transition 
could occur in two steps: first, the $\it bcc\/$ transforms into a partially ordered monoclinic phase driven by  rotational-translational coupling. Second, after a further temperature decrease, 
another transition to an orientationally ordered phase without 
change of the symmetry of the lattice sites takes place. 

In the present paper we analyze analytically the dynamic orientational order
in clusters of rigid octahedral molecules of type $AF_6$ to
determine the contribution of various interactions to the total potential and
the nature of the temperature-driven solid-solid transformation of plastic
clusters.  The symmetry of the molecules, $O_h$, and the molecular
sites in the cluster, $O_h$  or $C_{2h}$,  are taken into account explicitly in our analysis, based on 
the theory of orientationally-disordered crystals  \cite{SEI40,BRA72}. The site position of each 
molecule is determined by the Cartesian coordinates of its mass center at $A$. 
Throughout the paper we refer to ({\it i\/}) a space (laboratory) axis system $XYZ$, fixed at the cluster's 
center of mass; ({\it ii\/}) a nonrotating system $X'$$Y'$$Z'$ parallel to $XYZ$ but with its
origin translating with the molecular center of mass; ({\it iii\/})  a rotating, body-fixed system $xyz$. For  rigid molecules, these axes coincide with the principle axes of inertia.

The translation  of a molecule is separable as the
motion of the molecular center of mass in the $XYZ$ coordinates.
The $xyz$ orientation in the $XYZ$ system
is given with $\Omega \equiv (\theta,\phi,\psi)$, where $\theta$ and  $\phi$ are
the ordinary polar coordinates of the $z$ axis in the $XYZ$ system and $\psi$ 
is an angle in the $xy$-plane measuring the rotation  clockwise about the $z$ axis.

The molecular orientation in orientationally-disordered  high-{\it T\/} crystals has been presented in terms of symmetry-adapted rotator functions used by James and Keenan \cite{JAM59} to describe the 
orientational phase of methane with tetrahedral symmetry.

Let us consider an octahedral molecule $n$ in its initial orientation 
$\Omega_n$
where the molecular axes coincide with the laboratory system axes.
The orientational density distribution is expanded in terms of
spherical harmonics $Y^m_l(\Omega)$. The molecular symmetry requires that
$l=0,4,6,....$ and only certain linear combinations of $Y^m_l$ occur. For each
allowed $l$, we determine the molecular symmetry-adapted functions

\begin{equation}
S^\lambda_l(\Omega_n)=
\sum_{m=-l}^l
Y^m_l(\Omega_n)\alpha_l^{m\lambda}
\label{sfmolecule}
\end{equation}

\noindent
where the superscript $\lambda$ refers to the identity representation
of the cubic group $O_h$ and $\alpha_l^m$
are tabulated in Ref.\cite{BRA72}. A molecular form factor is defined for the 
allowed $l$=0,4,6,...as

\begin{equation}
g_l^{\lambda} =
\sum_{\nu=0}^{N_a}
S^\lambda_l[\Omega_{\nu}(n)]
\label{formfactor}
\end{equation}

\noindent
if the molecular axes coincide with the space axes;
 $N_a$ is the number of the atoms in a molecule; $\Omega_{\nu}(n)$ 
denotes the orientation of atom $\nu$ in the space-fixed system.  

Unlike bulk crystals,  the cubic structure of a free cluster is  broken
at the cluster surface. In what follows, we consider the symmetry of the 
volume molecules, defined as molecules having all the neighbors required by
a specific point group. We neglect the symmetry-breaking
associated with the surface molecules, which belong to a different group, if any.
The separate description of the surface and the volume  limits the analysis. In the closing remarks, 
we discuss a possible correction to this assumption.

The orientational changes at a cubic site are also expressed in terms of 
site-symmetry-adapted functions \cite{BRA72}:

\begin{equation}
S^\tau_l(\Omega)=
\sum_{m=-l}^l
Y^m_l(\Omega)\alpha_l^{m\tau},
\label{sfspace}
\end{equation}

\noindent
where the superscript $\tau$ = ($G$, $\Gamma$, $p$, $\rho$) indicates
the irreducible representations $\Gamma$ of the 
group $G$, $p$ distinguishes between the representations that occur more
than once, and $\rho$ denotes the rows of a given representation. The 
symmetry-adapted functions represent a complete basis in the 
$\Omega(\theta,\phi,\psi)$-space.
 The $l=4$ manifold of this system
reduces to the representation $A_{1g}$, $E_g$, $T_{1g}$, $T_{2g}$ under 
$G \equiv O_h$
and to the representations 5$A_g$ and 4$B_g$ under $G \equiv C_{2h}$. 
For the normalized
function $S^\tau_4$ we find:

$\alpha_4^{m\tau} = [0.763, m = 0; 0.457, m =\pm 4]$ for $O_h$, $A_{1g},1,1$

$\alpha_4^{m\tau} = [0.645, m = 0; -0.541, m =\pm 4]$ for $O_h$, $E_g,1,1$

$\alpha_4^{m\tau} = [-0.707, m =\pm 2]$ for $O_h$, $E_g,2,1$

$\alpha_4^{m\tau} = [-i0.663, m = \pm 1; -i0.25 , m =\pm 3]$ for $O_h$, 
$T_{1g},1,1$

$\alpha_4^{m\tau} = [\pm i0.663, m = \pm 1; -i0.25 , m =\pm 3]$ for $O_h$, 
$T_{1g},2,1$

$\alpha_4^{m\tau} = [\pm i0.707, m = \pm 4]$ for $O_h$, $T_{1g},3,1$

$\alpha_4^{m\tau} = [\pm i0.663, m = \pm 3; \pm i0.25 , m = \pm 1]$ 
for $O_h$, $T_{2g},1,1$

$\alpha_4^{m\tau} = [\pm 0.663, m= \pm 3; \pm 0.25 , m = \pm 1]$ 
for $O_h$, $T_{2g},2,1$

$\alpha_4^{m\tau} = [\pm i0.707, m = \pm 2]$ for $O_h$, $T_{2g},3,1$.

\noindent
The largest contribution to the crystal field is that of the $A_{1g}$
component of the $l = 4$ manifold.

At low temperatures, clusters of $AF_6$ molecules adopt a monoclinic structure
$C_{2h}$. 
All coefficients $\alpha^\tau$ are equal to 1 for
the five-fold representation $A_g$  $\left(m = 0,\pm 2,\pm 4\right)$ and
for the four-fold representation $B_g$ $\left(m = \pm 1, \pm 3\right)$.

An arbitrary molecular orientation $\Omega_{n'}$ with respect to the initial
one $\Omega_n$ is obtained by a rotation specified with the Euler angles
(${\bf \omega}$=$\alpha,\beta,\gamma)$.  Rotation $\Omega^b$ of a
molecule does not affect the spherical harmonics
$Y^k_l(\Omega^b)$ defined in the body system. In the space system 
these are changed to
$\hat{R}(\omega)Y^k_l(\Omega^b)=
\sum_{m=-l}^l
Y^m_l(\Omega^s)D^{mk}_l(\omega)$, where  $\Omega^s$ determines the space orientation of the molecule.
 For a molecule in arbitrary
orientation $\Omega_{n'}$, the symmetry-adapted function changes to

\begin {equation}
\hat{R}(\omega)S^\lambda_l(\Omega^b)=
\sum_{k=-l}^l
\sum_{m=-l}^l
Y^m_l(\Omega^s)
D^{mk}_l(\omega)
\alpha^{k\lambda}_l
\label{sinspace}
\end{equation}

\noindent
with $D^{mk}_l(\omega)$ the Wigner matrices.
We determine the spherical harmonics $Y^m_l(\Omega^s)$ from the 
equation inverse to (\ref{sfmolecule}) and  put it in
(\ref{sinspace}). The result is

\begin {equation}
\hat{R}(\omega)S^\lambda_l(\Omega^b)=
\sum_{\tau,m,k}S^\tau_l(\Omega^s)
\left(\alpha^{m\tau}_l\right)
D^{mk}_l(\omega)\alpha^{k\lambda}_l
\label{sbodytospace}
\end{equation}

The equation (\ref{sbodytospace}) relates the symmetry-adapted functions 
$S^\lambda_l(\Omega^b)$
for the body (molecular) system  and the symmetry-adapted functions
$S_l(\Omega^s)$
for the space (cluster) system. Rewriting the eq. (\ref{sbodytospace}) as
$\hat{R}(\omega)S^\lambda_l(\Omega^b)=
\sum_{\tau}
S^\tau_l(\Omega^s)
\Delta^{\tau\lambda}_l(\omega)$, where
$\Delta^{\tau\lambda}_l(\omega)=
\sum_{k=-l}^l
\sum_{m=-l}^l
\left(
\alpha^{m\tau}_l\right)
D^{mk}_l(\omega)
\alpha^{k\lambda}_l$  are the rotator functions
$\Delta^{\tau\lambda}_l(\omega)$ defined by the symmetry 
properties of the molecule $\alpha^{k\lambda}_l$ and of the site
$\alpha^{m\tau}_l$.  Rotator functions with
$l=3$ were introduced for solid $CD_4$ in \cite{JAM59}.
The rotator function's average value 
$\bar{\Delta}^{\tau\lambda}_l(\omega)$ 
is zero in the disordered phase and non-zero in the orientationally ordered
phase. This property makes it suitable to be chosen as an {\it order parameter\/}.

So far we have considered a single molecule at a specific site. The orientational
configuration of $N$ molecules in the cluster is given by
$\Delta^{\tau\lambda}_l(\omega(n)) = \Delta^{\tau\lambda}_l(n)$ where
$n=1,2,...,N$, labels
each molecule's center at its lattice position ${\bf r}_n$. The 
interaction between two molecules $n$ and $n'$ can be written as a sum
of atom-atom potentials \cite{BER97}:

\begin{equation}
V(n,n')=
\sum_{\nu,\nu'}^{N_a}
V(n,\nu;n',\nu')
\label{pair}
\end{equation}

\noindent
where $(n,\nu)$ labels
the $\nu^{th}$ atom in the molecule at site ${\bf r}_n$.
The total potential $V$ of $N$ molecules is:

\begin{equation}
V=
\sum_{n<n'}^N
\sum_{\nu,\nu'}^{N_a}
V(n,\nu;n',\nu')
\label{potential}
\end{equation}

The potential $V(n,\nu;n'\nu')$ depends on the distance $r_{\nu\nu'}$ 
between the atoms $\nu$ and $\nu'$. The position of the
$\nu^{th}$  atom in the $n^{th}$  molecule with respect to the space system
is given by

\begin{displaymath}
{\bf R}(n,d^\nu)= {\bf r}_n + d^\nu\Omega_\nu(n) + {\bf u}(n)
\end{displaymath}

\noindent
with ${\bf u}(n)$ being the displacement of the $n^{th}$ molecule from its 
site position ${\bf r}_n$.
$\Omega_\nu(n)$ indicates the orientation of the vector ${\bf r}_{n\nu}$ in the
space system and $d^\nu$ is its length. We
expand $V$ from Eq.(\ref{pair}) in terms of the displacements ${\bf u}(n)$:

\begin{equation}
V(n,n')=
\sum_{p=0}^\infty
\sum_{\nu,\nu'}
{1 \over(p!)}
V^{(p)}_{i_1...i_p}(r_{\nu\nu'})
\left[
u_{i_1}(n)-u_{i_1}(n')
\right]
...
\left[
u_{i_p}(n)-u_{i_p}(n')
\right]
\label{expand}
\end{equation}

\noindent
with the notation

\begin{equation}
V^{(p)}_{i_1...i_p}(r_{\nu\nu'}) =
{\partial^p V(r_{\nu\nu'}) \over \partial (r_{\nu\nu'})_{i_1}...
\partial (r_{\nu\nu'})_{i_p}}\mid_{u=0}
\label{partial}
\end{equation}

The coefficients $V^p$ contain the orientational dependence of the molecules
at the sites $n$ and $n'$. We expand them in terms of symmetry-adapted functions
$S^\tau_l$ (\ref{sfspace}). In the following we write $S_\mu(\nu)$ for
$S^\tau_\l(\Omega_\nu)$, where $\mu \equiv \mu(\tau, l)$:

\begin{equation}
V^p_{i_1...i_p}(r_{\nu\nu'}) =
\sum_{\mu\mu'}
c^{(p)}_{i_1...i_p\mu\mu'}(n,n')
S_\mu(\nu)S_{\mu'}(\nu').
\label{expandsf}
\end{equation}

The coefficients 
$c^{(p)}_{i_1...i_p\mu\mu'}(n,n')$
are determined from

\begin{equation}
c^{(p)}_{i_1...i_p\mu\mu'}(n,n') =
\int d\Omega_\nu
\int d\Omega_{\nu'}
V^{(p)}_{i_1...i_p}(n,n')
S_\mu(\nu)S_{\mu'}(\nu').
\label{coefc}
\end{equation}

We put $V^{(p)}$ from Eq.(\ref{expandsf}) in Eq.(\ref{expand}) and use the molecular form factor Eq.(\ref{formfactor}) if the molecular axes coincide
with the space axes or  $\sum_\nu$$S_\mu(\nu) = g_l^{\lambda}\Delta_\mu(\omega)$ if 
the molecule is rotated at an angle $\omega$. Thus the pair potential becomes:

\begin{equation}
V(n,n')=
\sum_p
\sum_{\mu\mu'}
{1 \over p!}
c^p_{i_1...i_p\mu\mu'}(n,n')
g_l^{\lambda}g_{l'}^{\lambda}
\Delta_\mu(n)
\Delta_{\mu'}(n')
\left[
u_{i_1}(n)-
u_{i_1}(n')\right]
...
\left[
u_{i_p}(n)-
u_{i_p}(n')\right]
\label{pairfinal}
\end{equation}

Molecular and site symmetry considerations restrict the number of terms in the
sums, thus reducing the computational effort to obtain the contribution of the
different interactions in the total pair potential Eq. (\ref{pairfinal}).

The value of $p=0$ corresponds to a rigid lattice (no displacements of the
molecular center of mass). For this case we get only 
rotational-rotational interaction between two molecules  with
$\mu\not= (0,0)$ and $\mu'\not= (0,0)$:

\begin{equation}
V^0(n,n')=
\sum_{\mu\mu'}
c^0_{\mu\mu'}(n,n')g_l^{\lambda}g_{l'}^{\lambda}
\Delta_\mu(n)
\Delta_{\mu'}(n')
\label{rotrot}
\end{equation}

The total rotational interaction is the sum over all molecules $\sum_{n,n'}^N{V^0(n,n')}$.

The matrix of rotational-rotational interaction is defined by

\begin{equation}
\hat{J}_{\mu\mu'}(n,n')=
c^{(0)}_{\mu\mu'}(n,n')g_l^{\lambda}g_{l'}^{\lambda}
\label{matrix}
\end{equation}

\noindent
where 

\begin{equation}
c^{(0)}_{\mu\mu'}(n,n')=
\int d\Omega_\nu
\int d\Omega_{\nu'}
V^{(0)}_{i_1...i_p}(n,n')
S_\mu(\nu)S_{\mu'}(\nu').
\label{coefc0}
\end{equation}

The structure of the interaction matrices $c^{(0)}_{\mu\mu'}(n,n')$ depends on the symmetry
of $S_\mu$ and on the relative position $(n,n')$  of  two
interacting molecules on a lattice with a symmetry specified by $\tau$.

The cluster transforms from a disordered cubic to an ordered monoclinic structure at $T_c$ that  is the temperature at which the free energies of the two forms are equal.  In order to calculate 
$T_c$ we also need the total field acting on the molecule at site ${\bf r}(n)$. The zero$^{th}$ 
approximation is to consider spherical-symmetrical molecules $\mu'=(0,0)$ acting 
on a molecule $n$
 $\mu\not=(0,0)$ on a rigid lattice $p=0$:  
$V^{(0)}(n,n')\mid_{l'=0}=
\sum_\mu
c_\mu^{(0)}(n,n')g_l^{\lambda}g^{\lambda}\Delta_\mu(n)$. 
Setting $\mu'=(0,0)$  
yields $S_{\mu'}$= $S^{A_{1g}}_0=(4\pi)^{-1/2}$ and $g_0= N_a(4\pi)^{-1/2}$. 
The coefficients $c^{(0)}_\mu(n,n')$ become:
$c^{(0)}_{\mu}(n,n') =
{1 \over \surd(4\pi)}
\int d\Omega_\nu
\int d\Omega_{\nu'}
V^{(0)}(n,n')
S_\mu(\nu)$.

Let us denote the interaction matrices $c^{(0)}_\mu(n,n')$ weighted with the
molecular factors $g_l^{\lambda}$ and $g_0^{\lambda}$
with $\upsilon^R_a$, where $a$ is an index for $(l,A_{1g},p,\rho)$: 
$\upsilon^R_a=
\sum_{n'}
c^{(0)}_a(n,n')
g_l^{\lambda}g_0^{\lambda}.
\label{upsilon}$. The crystal field acting on the molecule $n$ is:

\begin{equation}
V^R(n)=
\sum_a
\upsilon^R_a
\Delta_a(\omega,n)
\label{rotfield}
\end{equation}

\noindent

The rotator functions $\Delta^{A_{1g}}_l(\omega)$ in 
Eq.(\ref{rotfield}) are cubic functions.

The pair vibrational-rotational interaction is obtained 
from (\ref{pairfinal}) for  $p=1$, $\mu\not=(0,0)$, and
$\mu'=(0,0)$:

\begin{equation}
V^{TR}(n,n')=
V^{(1)}(n,n')=
\sum_{i\mu}
c^{(1)}_{i\mu}(n,n')
g_l^{\lambda}g_0^{\lambda}
\Delta_\mu(n)
\left[u_i(n) -
u_i(n')
\right].
\label{TR}
\end{equation}

The sum over all molecules results in the total bilinear interaction
$V^{TR} = 
\sum_{n,n'}V^{TR}(n,n')$.

The translational-orientational interaction is caused by the change of the orientational potential 
due to the displacement of the nearest neighbors.

The pair vibrational-vibrational interaction is obtained 
from Eq.(\ref{pairfinal}) for  $p=2$, $\mu'=(0,0)$, $\mu=(0,0)$:

\begin {equation}
V^{TT}(n,n')=
V^{(2)}(n,n')=
\sum_{i_1,i_2}
{1 \over 2}
c^{(2)}_{i_1i_2}(n,n')g_0^{\lambda}g_0^{\lambda}
\left[u_{i_1}(n)-
u_{i_1}(n')
\right]
\left[
u_{i_2}(n)-
u_{i_2}(n')
\right]
\label{TT}
\end{equation}

This gives for the total vibrational-vibrational interaction $V^{TT}=
\sum_{n,n'}V^{TT}(n,n')$. Now the total potential is:

\begin{equation}
V=
V^R+V^{TT}+V^{TR}+V^{RR}
\label{totalpot}
\end{equation}

\noindent
The equation (\ref{totalpot})  may be expanded with higher order terms which
may become important in some structural phase transitions \cite{GIN80}.

Having determined the interactions and the total field, we can calculate the
free energy $F$ of each phase as a function of the rotator functions considered
as order parameters \cite{LYN94}:

$F=
0.5\sum_q
\left[
\hat{1}\chi^{-1}_0
+
{\tt FT}[\hat{J}]
\right] 
\delta_\mu(q)\delta_{\mu'}(-q)$, 
where ${\tt FT}(\hat{J})$ and $\delta_\mu(q)$ are the Fourier images of the
rotator matrix $\hat{J}$ and $\Delta_\mu(\omega)$, respectively; 
$\hat{1}$ is the $3$x$3$ unit matrix;
$\chi_0 \equiv xT^{-1}$ is the single molecule orientational 
susceptibility \cite{MIC92}:
$x=Z^{-1}g_l
\int d\omega exp(-V^R(\omega)/T)
\left(
\Delta_\mu(\omega)\right)^2$ with $Z=\int d\omega exp(-V^R/T)$ the partition function.
The expectation value of $x$ does not depend on the components of the 
rotator function $\Delta$.  Two phases of clusters coexist in equal amounts 
or with equal frequency when their free energies are equal. In the 
limit of $N \rightarrow \infty$,
a phase transition occurs at $T_c$, which is the point where an eigenvalue 
of $[\hat{1}T +x(T){\tt FT}[\hat{J}]$ vanishes, if the transition is continuous. The temperature 
dependence of $x$ is very weak which means that the Curie-Weiss law $\chi_0=x(T)/(T-T_c)$ is 
valid for negative diagonal elements of $\hat{J}$. The transition point 
$T_c$ occurs at the largest value of the matrix for the representations allowed 
by the symmetry of the system, i.e.  $T_c=max[-x\hat{J}]$.

As an example, we have applied  this 
group-theoretical approach to the case of a cluster containing 59 $TeF_6$ molecules and have compared the results to those published in \cite{BER97,BER98,BER98A}.
In order to  account for the broken symmetry at the cluster surfaces, 
we use sum over the nearest neighbors. Since the molecules of type $AF_6$ have no low-order electrostatic moments, the Coulomb contribution to the Lennard-Jones potential can be neglected, 
see fig.2 in \cite {BER99}.  

We compare the rotational-rotational contributions, Eq.(\ref{rotrot}),
with their vibrational-rotational counterparts, (\ref{TR}),
for the $TeF_6$  molecule with the nearest neighbors located at
sites with cubic or monoclinic symmetry. For the purpose, we determine the vibrational spectrum,
the crystal field (\ref{rotfield}), and the rotational matrix (\ref{matrix}).The CERN Library \cite{CERN} is used to compute he elements of (\ref{matrix}). 
The contribution from $l=l'=4$ is the largest. The  $\hat{J}$ matrices in a 
cubic symmetry environment is:

$$\hat{J}_{T2g}=\left(\matrix{0.011&-0.0007&0\cr
-.0007&-0.02&0\cr
0&0&1.63\cr}\right)$$

$$\hat{J}_{T1g}=\left(\matrix{-0.004&-0.001&0\cr
-.001&-0.011&0\cr
0&0&-3.81\cr}\right)$$

$$\hat{J}_{Eg}=\left(\matrix{0.17&-0.02\cr
-0.02&2.41\cr}\right)$$.

These matrices are diagonal in $C_{2h}$ because it has only one-dimensional
representations. The diagonal elements of $\hat{J}$ in the case of
$C_{2h}$ are:

$\hat{J}(X_{11})=[0.043, 0.043, -0.027, -0.027, 0.014, 0.014, -0.011, -0.011]$

$\hat{J}(X_{12})=[-0.41, -0.41, 0.018, 0.018, -0.003, -0.003, -0.002, -0.002]$

$\hat{J}(X_{13})=[-0.22, -0.22, -0.08, -0.08, -0.0005, -0.0005, -0.0004, -0.0004]$

The largest value of $\hat{J}$ is obtained for the representation $A_{1g}$, 
so we choose as an order parameter
the rotator functions for this representation and expand the free 
energy $F$ in terms of the corresponding rotator functions. $A_{1g}$ is the only
common representation for $O_h$ and $C_{2h}$. To find correlation of $O_h$ and
$C_{2h}$ we carry out the correlation in two steps: first, pass from $O_h$ to
$D_{4h}$ and then imply the table for $D_{4h}$ to go on to $C_{2h}$ \cite{BRA72}.

The coefficients $\upsilon^R$ necessary to compute the crystal 
field (\ref{rotfield}) are calculated from  Eq.(\ref{coefc0}). We get for the
cubic symmetry $\upsilon^R_{O_h}=-7.98$ in the approximation of the nearest
eight neighbors. For the monoclinic structure, $C_{2h}$, this coefficient is
$\upsilon^R_{O_h}=-0.61$. From Eq.(\ref{formfactor}) we obtain for the
octahedral molecule of $TeF_6$: $g_0$=1.98 and $g_4$=1.29.

In the approximation
of the nearest neighbors interaction we have determined the energy per molecule: in $O_h$ , the rotational-rotational energy is 4$meV$ and the vibrational-rotational is 1$meV$; in $C_{2h}$, these values are 1.2$meV$ and 0.02$meV$, respectively. The conclusion is that the vibrational-rotational
interaction can be neglected in the ordering of molecules on monoclinic
sites, so that the lower-temperature transition is entirely driven
by rotational ordering and that transition is continuous even in small systems. However on cubic sites, motion of the molecular
centers of mass must be taken into account, and the transition acts
at least in small systems like it is first-order. The highly 
degenerate state of an octahedral molecule in the octahedral environment is 
resolved by a distortion of the cluster if the model requires a rigid molecule.
In this case we must deal with the Jahn-Teller effect that 
distorts the cluster. This implies that
a transition to a lower-symmetry structure is initiated. The larger value of the
rotational-rotational interaction, however, leads to a partial ordering of the
molecules which we clearly see \cite{BER98} in the thermodynamically 
less favored state  of clusters of
small size. The transition from a cubic to monoclinic structure is resolved
with the appearance of the representation of $E_g$ in $D_h$ which 
is equivalent to condensation of an active mode.

Taking into account the rotational-rotational interaction in the total field
 (\ref{totalpot}) for the orientational ordering in $C_{2h}$  we obtain $T_c={\it max} [-x\hat{J}] = 27K$. This is in a good agreement with 
the result ($\sim30K$) computed in the  molecular dynamics simulations 
\cite{BER97}. Thus we show 
that the choice of cubic rotator functions  (\ref{rotfield}) as an
order parameter is suitable for describing the mechanism of phase changes.

A final comment is that the theory of discrete point groups used in the present work sets limits on handling at once the surface and volume symmetries. However, the approach
of continuous symmetry measures as developed by Zabrodsky, Pinsky and 
Avnir \cite{AVI92,AVI98} might make it possible to
bring together the surface and the volume symmetry properties.

\bigskip

{\bf Acknowledgments}
The research was partially supported by Grant No.3270/1999 from the
Scientific Fund at the University of Sofia and Grant No. CHE-9725065 from the 
National Science Foundation.

\newpage


\newpage

\end{document}